\documentstyle[prl,aps,multicol,epsfig]{revtex}
\begin{document}
\draft

\renewcommand{\narrowtext}{\begin{multicols}{2}}
\renewcommand{\widetext}{\end{multicols}}
\newcommand{\be}{\begin{equation}}
\newcommand{\ee}{\end{equation}}
\newcommand{\lsim}   {\mathrel{\mathop{\kern 0pt \rlap
  {\raise.2ex\hbox{$<$}}}
  \lower.9ex\hbox{\kern-.190em $\sim$}}}
\newcommand{\gsim}   {\mathrel{\mathop{\kern 0pt \rlap
  {\raise.2ex\hbox{$>$}}}
  \lower.9ex\hbox{\kern-.190em $\sim$}}}
\def\be{\begin{equation}}
\def\ee{\end{equation}}
\def\ba{\begin{eqnarray}}
\def\ea{\end{eqnarray}}
\def\d{{\rm d}}
\def\i{{\rm i}}
\def\e{{\rm e}}
\def\ap{\approx}
\def\stot{\sigma_{\rm tot}}
\def\sel{\sigma_{\rm el}}
\def\skk{\sigma_{N\nu}^{\rm KK}}
\def\Ms{M_{\rm st}}
\def\Mp{M_{\rm Pl}}

\title{\hfill \begin{minipage}{4cm}
              {\small CERN-TH 2000-134}\\
              {\small UPR-0883-T}
              \end{minipage}
       \vskip1.0cm
       Ultrahigh energy neutrino interactions and weak-scale string theories}

\author{M.~Kachelrie{\ss}$^1$ and M.~Pl\"umacher$^2$}
\address{$^1$TH Division, CERN, CH-1211 Geneva 23, Switzerland}
\address{$^2$Department of Physics and Astronomy, University of Pennsylvania,
  Philadelphia, Pennsylvania 19104, USA}

\date{\today}

\maketitle

\begin{abstract}
It has been suggested that ultrahigh energy neutrinos can acquire
cross-sections approaching hadronic size if the string scale is as low
as $1-10$~TeV. In this case, the vertical air showers observed with
energies above the Greisen-Zatsepin-Kuzmin cutoff at $E\ap 6\cdot
10^{19}$~eV could be initiated by neutrinos which are the only known
primaries able to travel long distances unimpeded.  We have calculated
the neutrino-nucleon cross-section $\skk$ due to the
exchange of Kaluza-Klein excitations of the graviton in a field
theoretical framework. We have found that $\skk$ and the
transferred energy per interaction are too small to explain vertical
showers even in the most optimistic scenario.
\end{abstract}


\narrowtext

\section{Introduction}

Several experiments using different techniques have observed ultrahigh
energy cosmic rays (UHE CR) with energies up to $3\cdot
10^{20}$eV~\cite{rev,exp}. The isotropy of the UHE CR arrival
directions argues for their extragalactic origin, since galactic
and extragalactic magnetic fields cannot isotropize charged particles
of such energies. However, all known extragalactic sources of UHE CR,
such as AGN~\cite{bi}, topological defects~\cite{td} or the Local
Supercluster~\cite{bg}, result in a well pronounced
Greisen-Zatsepin-Kuzmin (GZK) cutoff~\cite{GZK} at $E_{\rm GZK}\ap
6\cdot 10^{19}$~eV, although in some cases the cutoff energy is
shifted closer to $1\cdot 10^{20}$~eV \cite{bg}. 

As possible solution to this puzzle it has been proposed that the UHE
primaries initiating the observed air showers are not protons, nuclei
or photons but neutrinos~\cite{alt,do98,ja00}. Neutrinos are the
only known stable particles which can traverse extragalactic space
even at energies $E\gsim E_{\rm GZK}$ without attenuation, thus
avoiding the GZK cutoff.

Although only two dozens of UHE events have been observed, we can pin
down rather precisely the required interactions of UHE neutrinos. 
Since the shower profile of Fly's Eyes highest energy event with 
$E\ap 3\cdot 10^{20}$~eV is well fitted by a proton \cite{p} and also
the lateral electron and muon distributions observed by AGASA are
consistent with this hypothesis, neutrino-nucleon interactions should 
mimic
nucleon-nucleon interactions at cms energies $\sqrt{s}\ap 500$~TeV.  
In particular, the neutrino-nucleon cross-section should reach
$\sigma=100-200$~mbarn, while the average energy fraction $y$
transferred per interaction to the shower should be large, $y\ap 0.6$. 

Most models introducing new physics at a scale $M$ to produce large
cross-sections for UHE neutrinos fail because experiments generally
constrain $M$ to be larger than the weak scale, $M\gsim m_Z$, and
unitarity limits cross-sections to be ${\cal O}(\stot)\lsim 1/M^2
\lsim 1/m_Z^2$~\cite{bu98}.  String theories with large extra
dimensions \cite{ex} are different in this respect: if the Standard
Model (SM) particles are confined to the usual $3+1$-dimensional space and
only gravity propagates in the higher-dimensional space, the
compactification radius $R$ of the large extra dimensions can be
large, corresponding to a small scale $1/R$ of new physics. The
weakness of gravitational interactions is a consequence of the large
compactification radius, since Newton's constant is then given by
$G_N^{-1}=8\pi R^{\delta} M_D^{\delta+2}$, where $\delta$ is the
number of extra dimensions and $M_D\sim\;$TeV is the fundamental mass
scale.  Such a scenario is naturally realized in theories of open
strings \cite{type1}, where SM particles correspond to open strings
beginning and ending on D-branes, whereas gravitons correspond to
closed strings which can propagate in the higher-dimensional space.
From a four-dimensional point of view the higher dimensional graviton
in these theories appears as an infinite tower of Kaluza-Klein (KK)
excitations with mass squared $m_{\vec{n}}^2=\vec{n}^2/R^2$.  Since
the weakness of the gravitational interaction is partially compensated
by the large number of KK states and cross-sections of reactions
mediated by spin 2 particles are increasing rapidly with energy, it
has been argued in Refs.~\cite{do98,ja00} that neutrinos could
initiate the observed vertical showers at the highest energies.

In the calculations of Refs.~\cite{ja00,nu99} it was assumed that the 
massless four-dimensional graviton and its massive KK excitations couple 
with the usual gravitational strength $\overline{M}_{\mbox{\small
Pl}}^{-1}= \sqrt{8\pi}/M_{\mbox{\small Pl}}$. Then the sum over all KK
contributions to a given scattering amplitude only converges in the
case of one extra dimension, and for two or more extra dimensions a
cutoff has to be introduced by hand. However, it has recently been
pointed out \cite{fluct} that due to brane fluctuations the effective
coupling $g_{\vec{n}}$ of the level $\vec{n}$ KK mode to
four-dimensional fields is suppressed exponentially,
\begin{equation}
  g_{\vec{n}}={1\over\overline{M}_{\mbox{\small Pl}}}
  \exp\left(-{c\,m_{\vec{n}}^2\over \Ms^2}\right)\;,
  \label{coupling}
\end{equation}
where $c$ is a constant of order $1$ or larger, which parametrizes the
effects of a finite brane tension \cite{fluct}, and $\Ms$ is the
string scale.  This exponential suppression thereby provides a
dynamical cutoff in the sum over all KK modes.  We have recalculated
the neutrino-nucleon cross-section $\skk$ due to the exchange of KK
gravitons in a four-dimensional, effective field theory valid for
$s\lsim\Ms^2$, taking this dynamical cutoff into account.  As a
consequence, our result for $\skk$ is considerably smaller than in
previous calculations~\cite{nu99}. Since we have found moreover that
the energy transfer per interaction is small at the
energies of interest, $y\ap 0.1$,  neutrinos
behave like deeply penetrating particles and cannot explain the
observed vertical air showers. This conclusion holds even if one
extrapolates the unitarity violating cross-section
valid for $s\lsim\Ms^2$ to the region $s\gsim\Ms^2$.

We have also derived an upper bound for $\skk$ consistent with unitarity 
for $s\gsim\Ms^2$ using the eikonal method. In this case, the resulting 
cross-section respects the Froissart bound and is numerically too
small to lead to observable consequences in UHE CR experiments.

\section{Neutrino-nucleon interaction via exchange of Kaluza-Klein gravitons}

\subsection{Cross-section in the low-energy limit, $s\ll\Ms^2$}

At energies below the string scale, effects from KK excitations can be
taken into account in an effective four-dimensional field theory. In
our calculation we have used the Feynman rules derived in
Ref.~\cite{feynman}, replacing the KK coupling
$\overline{M}_{\mbox{\small Pl}}^{-1}$ by the suppressed coupling
given in Eq.~(\ref{coupling}). It is then straightforward to compute
the gravitational contributions to partonic cross sections for
quark-neutrino scattering
\ba
 \frac{\d\sigma}{\d x \d\hat t} & = & \frac{1}{512\pi\hat s^2}
                                      \frac{P^2(\hat t)}{M_D^{4+2\delta}}
\nonumber\\ & \times & 
  \left[ 32\hat s^4 + 64 \hat s^3 \hat t 
       + 42\hat s^2 \hat t^2 + 10\hat s \hat t^3 + \hat t^4 
  \right] \,,  
  \label{nuquark}
\ea
and for gluon-neutrino scattering
\be
 \frac{\d\sigma}{\d x \d\hat t} = \frac{1}{32\pi\hat s^2}
                             \frac{P^2(\hat t)}{M_D^{4+2\delta}} 
                             \left[ 2\hat s^4 + 4 \hat s^3 \hat t 
                                    + 3\hat s^2 \hat t^2 + \hat s\hat t^3 
                              \right] \,. 
  \label{nugluon}
\ee
Here we have introduced the Bjorken variable $x$ and $\hat s=xs$,
$\hat t=xt$, where $s=2m_N E_\nu$ is the squared cms energy and $t$ 
the invariant momentum transfer. The terms in the brackets are
symmetric under the exchange $s\leftrightarrow u$ and agree therefore
with those of Ref.~\cite{ja00}. The function $P({\hat t})$ denotes the sum
over the propagators of the KK modes, including the couplings from
Eq.~(\ref{coupling}):
\be   \label{P}
  P({\hat t})=R^{-\delta}\sum\limits_{\vec{n}}
    {\exp\left(-{cm_{\vec{n}}^2\over \Ms^2}\right)\over 
    {\hat t}-m_{\vec{n}}^2}\;.
\ee
Due to the small separation $\sim1/R$ between the KK levels, the sum
can be approximated by an integral and $P({\hat t})$ is given by
\be
 P({\hat t}) =-\pi^{\delta/2}(-{\hat t})^{\delta/2-1}
    \exp\left(-{c{\hat t}\over \Ms^2}\right)
    \Gamma\left(1-{\delta\over2},-{c{\hat t}\over \Ms^2}\right) \,,
\ee
where $\Gamma(a,x)$ is the incomplete gamma function as defined, e.g.,
in Ref.~\cite{erd}. 
The $D=4+\delta$ dimensional mass scale $M_D$ and the string scale
$\Ms$ are expected to be of the same order and we have set them equal
in the following. Further, in our numerical examples we always
consider the case of two extra dimensions, $\delta=2$. However, our
results also hold for $\delta>2$.

\begin{figure}
\begin{picture}(8,8.5)
 \put(2.5,1.0) { \epsfig{file=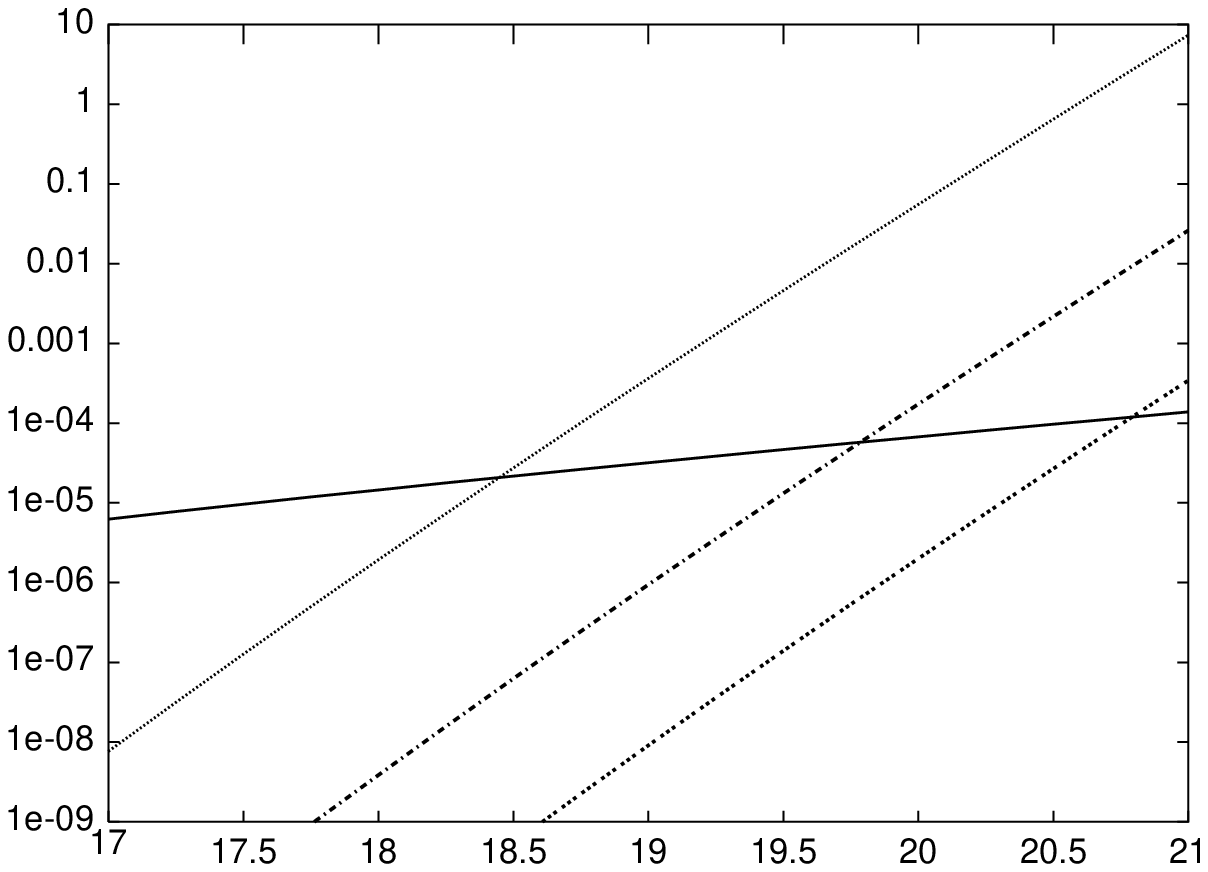,height=7.cm,width=10.cm,angle=0} }
 \put(7.0,0.6) {$\log(E_\nu/{\rm eV})$}
 \put(1.9,4.5) {$\sigma_{\nu N}$}
 \put(4.5,4.1) {CC}
 \put(5.9,3.2) {6 TeV}
 \put(8.3,3.2) {15 TeV}
 \put(10.3,3.2) {30 TeV}
\end{picture}
\caption{Neutrino-nucleon cross-section $\sigma_{\nu N}$/mbarn due to
  $W$-exchange (CC) and exchange of KK gravitons as function
  of $\log(E_\nu/{\rm eV})$ for $\Ms=6,15$ and 30~TeV. All for
  $\delta=2$ and $c=1$.  
\label{fig_sigma}}
\end{figure}

We have used the CTEQ4-DIS~\cite{cteq4} parton distribution functions (pdf) 
to calculate the total nucleon-neutrino cross-section. 
In Fig.~\ref{fig_sigma}, the  cross-sections
due to KK exchange are plotted for three different value of $\Ms$.
For comparison, we show also the charged-current cross-section of the SM.
We have neglected the neutral-current contribution because it is not
much larger than the uncertainty of the pdf's.
From Fig.~\ref{fig_sigma} it is clear that even for $\skk=1-10$~mbarn
a value of $\Ms$ not much above 1~TeV is required. While present collider
experiments do not exclude this possibility, SN 1987A gives $M_D\gsim 50$~TeV
\cite{bounds}. Although the latter limit was obtained for a rather
conservative choice of supernova parameters, the astrophysical
uncertainties inherent in this bound make it plausible that
$M_D\sim 10$~TeV is still compatible with SN 1987A.

The second important quantity characterizing the development of an air
shower besides $\stot$  is the energy transfer 
\mbox{$y=(E_\nu-E_\nu^\prime)/E_\nu$.}  
In contrast to charged-current scattering where the
electromagnetic shower initiated by the charged lepton is practically
indistinguishable from a hadronic shower, only the hit nucleon can
initiate an air shower in KK scattering. Therefore, even a neutrino with
large $\stot$ will behave like a penetrating particle if it does not
transfer a large fraction of its energy per interaction to the
shower.

\begin{figure}
\begin{picture}(8,8.5)
 \put(2.5,1.0) { \epsfig{file=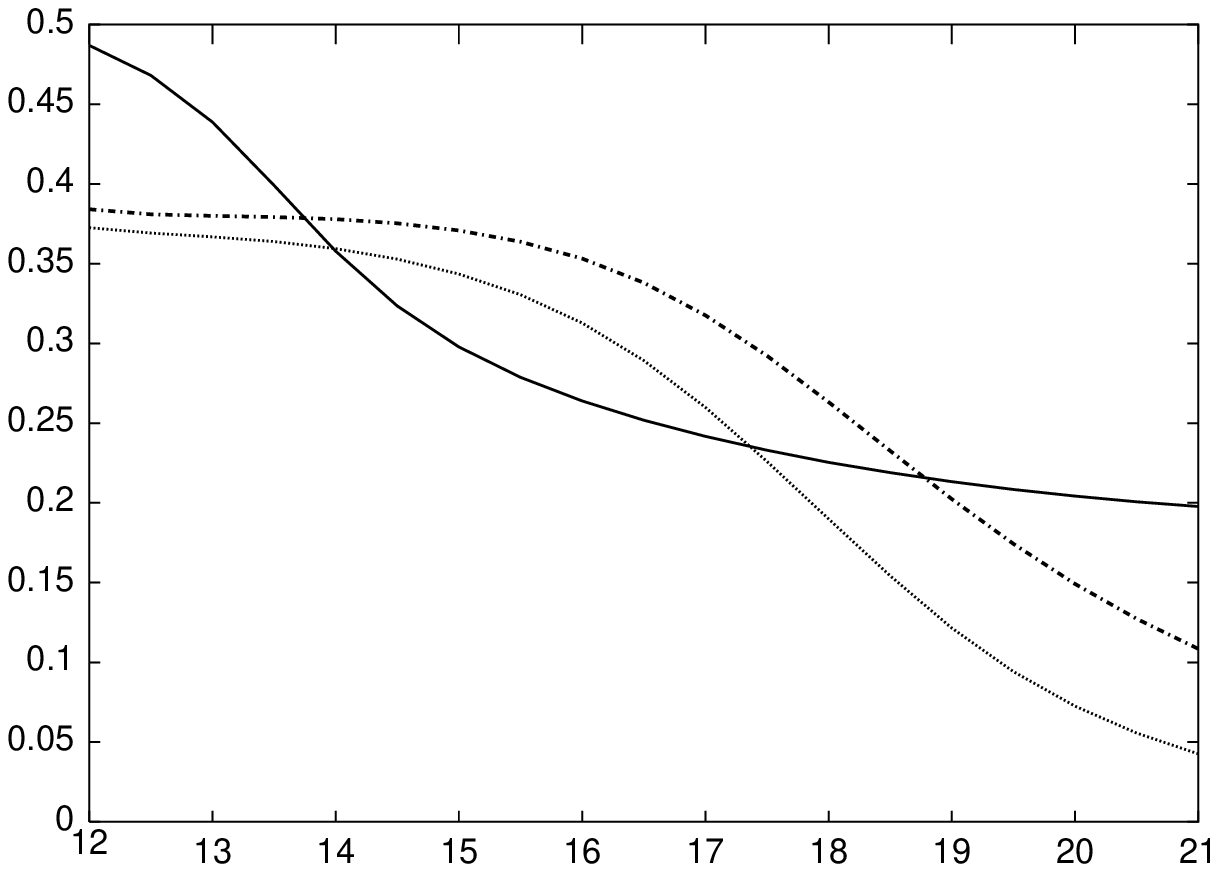,height=7.cm,width=10.cm,angle=0} }
 \put(7.0,0.6) {$\log(E_\nu/{\rm eV})$}
 \put(2.2,4.5) {$y$}
 \put(5.8,4.8) {CC}
 \put(10.8,3.2) {$g$}
 \put(10.8,2.2) {$q$}
\end{picture}
\caption{Energy transfer $y$ in the subreactions with $W$-exchange
  (CC), exchange of KK gravitons with quarks ($q$)
  and gluons ($g$) as function of $\log(E_\nu/{\rm eV})$ for
  $\Ms=6$~TeV, $\delta=2$ and $c=1$.
\label{fig_y}}
\end{figure}

In Fig.~\ref{fig_y}, the energy transfer $y$ is shown as function of
$E_\nu$. 
At energies of interest, $E_\nu\ap 10^{20}$~eV, the transferred energy
fraction is only around $y\ap0.1$, i.e., much smaller than $y\ap 0.6$
typical for nucleon-nucleon collisions.

\subsection{Cross-section in the high-energy limit, $s\gg\Ms^2$}

For energies comparable to the string scale $\Ms$, the effective
theory used above to derive $\skk$ breaks down. Since a calculation of
$\skk$ valid for $s\gg\Ms^2$ within string theory is beyond the scope
of this paper, we restrict ourselves in the following to obtain an
upper bound for $\skk$.  In deriving this bound, we will rely on the
assumption that string theory has a better high-energy behavior than
four-dimensional field theory. More concretely, we assume that
cross-sections do not grow faster with $s$ in string theory than
allowed by the unitarity bound derived in four-dimensional field
theory.

Let us use the Regge language~\cite{ed67,co77} to discuss the
high-energy behavior of the total cross-section $\stot$. A general
Regge amplitude $A_R$ can be represented by
\be
 A_R(s,t)= \beta(t) s^{\alpha(t)} \,,
\ee
where the exponent $\alpha(t)$ is given by the relation between spin
$\sigma_i={\rm int}[\alpha(t)]$ and mass $m^2_i=t$ of the particles
lying on the leading Regge trajectory contributing to the
reaction. In our case, the intercept $\alpha(0)$ of this 
trajectory is equal to the spin of the massless graviton, $\alpha(0)=2$. 
Since $\stot\sim s^{-1}\Im\{A(s,0)\}$, any Regge amplitude with
$\alpha(0)>1$ violates both unitarity and the Froissart bound
$\stot(s)<{\rm const.} \ln^2(s/s_0)$. 
In the case of the Pomeron with intercept $\alpha(0)\ap 1.1$, it is
well-known that this bad high-energy behavior can only be cured if
unitarization produces strong canceling cuts additionally to the
Regge poles. 

A convenient way to ensure unitarity is the use of the eikonal method.  
There, the amplitude $A(s,t)$ is given by~\cite{co77}
\be
 A(s,t)= 8\pi s \int_0^\infty \d b \;b  A_H(s,b)
 J_0(b\sqrt{-t})  \,,
\ee
where 
\be
 A_H(s,b)=\frac{\e^{\i\chi(s,b)}-1}{2\i}
\ee
and the eikonal function $\chi(s,b)$ is
\be    \label{chi1}
 \chi(s,b)=\frac{1}{8\pi s}\int_{-\infty}^0 \d t \; A_R(s,t)
 J_0(b\sqrt{-t}) \,.
\ee

We assume as Ref.~\cite{nu99} that the Regge trajectories 
are linear, $\alpha(t)=\alpha_0+\alpha't$, and that their slope is
given by the string tension, $\alpha'=1/(4\pi\Ms^2)$. The residue
\be 
 \beta(t)= -\exp(-\i\alpha(t)\pi/2) \: \e^{at} 
\ee
contains the phase of the amplitude and the Reggeon coupling 
$\propto\exp(at)$, for which Eq.~(\ref{coupling}) suggests
$a=c/\Ms^2$. Inserting $ A_R(s,t)$ into Eq.~(\ref{chi1}), we obtain
\ba  \label{chi2}
 \chi(s,b) & = & -\frac{(s/s_0)^{\alpha_0}\e^{-\i\alpha_0\pi/2}}{8\pi s d}\:
                \, \exp\left(-\frac{b^2}{4d}\right) =
\\
  & = & -\frac{\e^{-\i\alpha_0\pi/2}}{8\pi s_0 d}\:
                \, \exp\left[-\frac{b^2}{4d}+(\alpha_0-1)\ln(s/s_0)\right] \,,
\nonumber
\ea
where $d=a+\alpha'[\ln(s/s_0)-\i\pi/2]$. If $s\gg s_0$ and
\be
 b^2\gg b_0^2(s) = 4(\alpha_0-1) \alpha'\ln^2(s/s_0) \,,
\ee
then $\chi(s,b)\ap 0$ and $A_H(s,b)\ap 0$. In the opposite case,
$b^2\ll b_0^2$, the imaginary part $\Im[\chi(s,b)]\to\infty$ and
$A_H(s,b)=\i/2$.

The amplitude
\be
 A_H(s,b)= \left\{ \begin{array}{c} \i/2, \quad b^2\ll  b_0^2(s)\\
                                    0,              \quad b^2\gg  b_0^2(s)
                   \end{array} \right.
\ee
corresponds to complete absorption on a black disc with radius $b_0$.
Using $J_0(0)=1$ and the optical theorem, the energy dependence of the
total cross-section follows as
\ba   
 \stot(s) & = & 
   8\pi \int_0^\infty \d b \;b  \Im\{A_H(s,b)\} \ap 2\pi b_0^2(s) =
\nonumber\\  \label{Fr}
   & = & {\rm const.} \ln^2(s/s_0) \,.
\ea
Thus, our result respects the Froissart bound in contrast to the
corresponding results of Ref.~\cite{nu99}. The authors of \cite{nu99}
argued that the asymptotic behavior $\sigma(s)\propto s$ is natural to
expect because of the massless graviton. There are two arguments against 
this interpretation:  
First, the contribution of any individual KK mode to $\stot$ is
negligible. Therefore, we can omit in the summation over the KK modes
the massless graviton, i.e., we can omit the 
$n=0$ mode in Eq.~(\ref{P}). Then, there is a small,
but finite mass gap and the Froissart bound should hold.
Second, physical quantities like total cross-sections are
infrared-finite and although infrared divergences make formally 
the application of the Froissart bound impossible this should be regarded  
merely as a technical obstacle.

In Eq.~(\ref{Fr}), neither the scale $s_0$ nor the constant can be fixed
within the eikonal method. However, we can obtain an upper bound for
$\skk$ if we choose the constant as $\skk(s')$ and identify
$s'$ with the scale above which $s$ wave unitarity is
violated. An explicit calculation shows that, as expected, $s'$
coincides approximately with $\Ms^2$. Taking into account that the
number of possible targets grows in the nucleon like
$(s/s')^{0.363}$~\cite{cteq4},  
the total cross-section of neutrino-nucleon scattering due to
exchange of KK gravitons is bounded by
\be
 \stot(s) = \skk(\Ms^2) \ln^2\left(s/\Ms^2\right) 
            \left(s/\Ms^2\right)^{0.363}
 \,, \quad s\gsim \Ms^2  \,.
\ee

Finally, we stress that also in the derivation of $\stot$ in the Regge
picture the exponential suppression of high-lying KK modes was essential.

\section{Air showers}

First, let us discuss in a very general way how large the total
cross-section of an UHE primary able to produce the observed vertical air
showers should be. The survival probability $N$ at atmospheric depth
$X$ of a primary with mean free path $\lambda=m_{\rm air}/\stot$ 
is $N(X)=\exp(-X/\lambda)$, where $m_{\rm air}\ap 2.4\cdot 10^{-24}$~g
is the weight of an ``average'' air atom. Hence, the probability
distribution $p$ of the first interaction point $X_1$ has its maximum
at $p(X_1)=\lambda$. 

For a proton with energy $E=10^{20}$~eV, the mean free path is
$\lambda_p\ap 40$g/cm$^2$ and thus a proton air shower is indeed
initiated in the top of the atmosphere. After the first interaction,
the number of particles in the shower growths until it reaches its maximum at 
$X_{\rm max}\ap 800$~g/cm$^2$. Hence, a vertical proton air shower needs
almost the complete atmosphere for its development.

How would this picture change for a neutrino with $\lambda_\nu=10\lambda_p$,
i.e., $\stot=15$~mbarn? Taking into account only the delayed start of 
the shower shifts the shower maximum already $\ap 360$~g/cm$^2$ downwards in 
the atmosphere. The small energy fraction transferred to the shower per
interaction delays the shower development even further.
Additionally, the fluctuations of a neutrino shower are
enhanced compared to a proton shower. Hence, the shower evolution
is clearly different compared to a proton shower. 
In contrast to Ref.~\cite{ja00}, we conclude therefore that even 
neutrino-nucleon cross-sections as large as 15~mbarn due to KK
exchange are not sufficient to explain vertical air showers by
neutrino primaries.

If $\skk$ would be considerably larger than 15~mbarn, in principle a
detailed simulation of the neutrino air shower development would be
necessary. Such issues were studied in the case of glueballinos
$\tilde G$ \cite{be00}, where it was shown that $\tilde G$
showers are clearly distinguishable from proton showers even for
$\sigma_{N\tilde G}\ap 90$~mbarn and $y\ap 0.1$.
It is therefore very likely 
that the small energy transfer is sufficient to differentiate between
showers initiated by proton and by neutrinos interacting through KK
gravitons. 

Let us now briefly discuss the issue of horizontal air showers. The Fly's Eye
experiment presented an upper limit for the neutrino flux from the
non-observation of horizontal air showers~\cite{FE}. This yields a limit 
on $\Ms$ which is not competitive with accelerator bounds at present. The 
exact sensitivity of future experiments like AUGER or OWL is hard to estimate 
due to the unknown UHE neutrino flux.
Signatures would be the anomalous energy and zenith angle distribution 
of the neutrino showers. 

Finally, we address the question if weak-scale string theories can
offer additional signatures for UHE neutrino detection. Gauge bosons
and higgses could have KK towers of excitations similar to the
graviton \cite{anton}. In this case, the KK excitations of the $W^{\pm}$ 
boson with
mass $m_n^2=m_W^2+\vec n^2/R^2$ result in a corresponding tower of
Glashow resonances $\bar\nu_e+e^-\to W^-_n\to $ all.  Experimental
constraints from existing colliders limit the size of the extra
dimensions to be $R^{-1}\gsim1\;$TeV \cite{bounds}. Therefore, even
the first KK resonance at $E_{\rm res}\ap 1/(2m_e R^2)\ap
10^{18}$~eV$\;(R^{-1}/{\rm TeV})^2$ has a cross-section too small to
be distinguishable from the SM background.  Note also that the
couplings of KK excitations with $n>1$ are again exponentially
suppressed \cite{gauge}. Alternative suggestions like $s$-channel
exchange of leptoquarks \cite{leptoq} or squarks in supersymmetric
models with $R$-parity violation \cite{Rp} also fail to generate cross
sections of the required magnitude \cite{bu98}, even if possibly
existing KK-excitations of these states are taken into account. 

\section{Conclusions}

We have calculated the neutrino-nucleon cross-section $\skk$ due to the
exchange of KK excitations of the graviton taking into
account the exponential suppression of modes with $m_{\vec n}^2\gsim
\Ms^2$. Because of the smallness of the resulting cross-section and energy
transfer per interaction, the neutrino behaves also in these theories  
as a deeply penetrating particle. In the case that the cross-section
$\skk\propto s^2$ continues to grow for $s\gsim\Ms^2$, thereby
violating four-dimensional unitarity, future UHE CR experiments like
AUGER or OWL could be more sensitive to large extra dimensions than LHC.
However, an accurate determination of the sensitivity of these
experiments would require a string-theoretical calculation of the
neutrino-nucleon cross section.

\acknowledgements

We would like to thank Karim Benakli and Sergey Ostapchenko for
helpful comments and the Max Planck Institut f\"ur
Physik, where this work was started, for hospitality. 
M.P.\ was supported in part by the U.S.\ Department
of Energy Grant No.~EY-76-02-3071 and in part by the Feodor Lynen
Program of the Alexander von Humboldt Foundation.

\widetext
\newpage
\unitlength1.0cm
\quad\vskip1.0cm

\end{document}